\documentclass{article}

\usepackage[T1]{fontenc}
\usepackage{fullpage}

\DeclareUnicodeCharacter{03BB}{$\lambda$}
\DeclareUnicodeCharacter{2200}{$\forall$}
\DeclareUnicodeCharacter{2113}{$\ell$}

\usepackage{hyperref}

\usepackage{color}

\input{depfold.sty}

\title{Towards an induction principle for nested~data~types}
\author{Peng Fu and Peter Selinger}
\date{Dalhousie University}

\begin{document}

\maketitle 

\begin{abstract}
A well-known problem in the theory of dependent types is how to handle so-called \textit{nested data types}. These data types are
difficult to program and
to reason about in total dependently typed languages such as Agda and Coq. In particular, it is not easy to derive a canonical induction principle
for such types. Working towards a solution to this problem, we introduce \textit{dependently typed folds} for nested data types. Using the nested
data type $\Bush$ as a guiding example, we show how to derive its dependently typed fold and induction principle. We also discuss the
relationship between dependently typed folds and the more traditional higher-order folds.
\end{abstract}

\section{Introduction}
\label{sec:intro}
Consider the following list data type and its fold function in
Agda {\cite{Agda}}.
{\small\codecheck{77}
\begin{verbatim}
data List (a : Set) : Set where
  nil : List a
  cons : a -> List a -> List a

foldList : ∀ {a p : Set} -> p -> (a -> p -> p) -> List a -> p
foldList base step nil = base
foldList base step (cons x xs) = step x (foldList base step xs)
\end{verbatim}}

\noindent The keyword $\Set$ is a kind that classifies types.
The function $\foldList$ has two implicitly quantified type variables
$\texttt{a}$ and $\texttt{p}$. In Agda, implicit arguments are
indicated by braces (e.g., $\verb!{a}!$), and can be omitted.

The function $\foldList$ is defined by structural recursion and is
therefore terminating. Agda's termination checker automatically checks
this. Once $\foldList$ is defined, we can use it to define other
terminating functions such as the following $\mapList$ and $\sumList$. This
is similar to using the iterator to define terminating arithmetic
functions in System $\textbf{T}$ \cite[\S 7]{girard1989proofs}.

{\small\codecheck{87}
\begin{verbatim}
mapList : ∀ {a b : Set} -> (a -> b) -> List a -> List b
mapList f ℓ = foldList nil (λ a r -> cons (f a) r) ℓ

sumList : List Nat -> Nat
sumList ℓ = foldList zero (λ x r -> add x r) ℓ
\end{verbatim}
 }

 When defining the $\mapList$ function, if the input list is empty,
 then we just return $\nil$, so the first argument for $\foldList$ is
 $\nil$. If the input list is of the form $\texttt{\cons\ a as}$, we
 return $\texttt{\cons\ (f a) (\mapList\ f as)}$, so the second
 argument for $\foldList$ is $\texttt{(λ a r -> \cons\ (f a) r)}$,
 where $\texttt{r}$ represents the result of the recursive call
 $\texttt{\mapList\ f as}$. The function $\sumList$ is defined
 similarly, assuming a natural numbers type $\Nat$ with zero and
 addition.

We can generalize the type of $\foldList$ to obtain the following induction principle for lists.
\pagebreak
{\small\codecheck{95}
\begin{verbatim}
indList : ∀ {a : Set} {p : List a -> Set} ->
          (base : p nil) ->
          (step : (x : a) -> (xs : List a) -> p xs -> p (cons x xs)) ->
          (ℓ : List a) -> p ℓ
indList base step nil = base
indList base step (cons x xs) = step x xs (indList base step xs)
\end{verbatim}}
We can see that the definition of $\indList$ is almost the same as
that of $\foldList$. Compared to the type of $\foldList$, the type of
$\indList$ is more general as the kind of $\texttt{p}$ is generalized
from $\Set$ to $\texttt{\List\ a -> \Set}$. We call $\texttt{p}$ a
\textit{property} of lists. The induction principle $\indList$ states
that to prove a property $\texttt{p}$ for all lists, one must first
prove that $\nil$ has the property $\texttt{p}$, and then assuming that
$\texttt{p}$ holds for any list $\texttt{xs}$ as the induction
hypothesis, prove that $\texttt{p}$ holds for $\texttt{\cons\ x xs}$
for any $\texttt{x}$.

We can now use the induction principle $\indList$ to prove that
$\mapList$ has the same behavior as the usual recursively defined
$\mapListprime$ function.

{\small\codecheck{104}
\begin{verbatim}
mapList' : ∀ {a b : Set} -> (a -> b) -> List a -> List b
mapList' f nil = nil
mapList' f (cons x xs) = cons (f x) (mapList' f xs)

lemma-mapList : ∀ {a b : Set} -> (f : a -> b) -> (ℓ : List a) ->
                mapList f ℓ == mapList' f ℓ
lemma-mapList f ℓ =
  indList {p = λ y -> mapList f y == mapList' f y} refl
    (λ x xs ih -> cong (cons (f x)) ih) ℓ
\end{verbatim}}

In the proof of $\texttt{lemma-mapList}$, we use $\texttt{refl}$ to construct a
proof by reflexivity and $\texttt{cong}$ to construct a proof by congruence.
The latter is defined such that $\texttt{cong f}$ is a proof of
$\texttt{x == y -> f x == f y}$.
The key to using the induction principle $\indList$ is to specify
which property of lists we want to prove. In this case the property is
$\texttt{(λ y -> \mapList\ f y == \mapListprime\ f y)}$.

To summarize, the fold functions for \textit{ordinary} data types (i.e., non-nested inductive data types such as $\List$ and $\Nat$) are well-behaved in the following sense. (1)
The fold functions are defined by well-founded recursion. (2) The fold functions can be used to define a range of terminating functions (including maps). (3) The types of the fold functions
can be generalized to the corresponding induction principles.

Nested data types \cite{bird1998nested} are a class of data types that
one can define in most functional programming languages (OCaml,
Haskell, Agda). They were initially studied by Bird and Meertens
\cite{bird1998nested}. They have since been used to represent de
Bruijn notation for lambda terms \cite{bird1999bruijn}, and to give an
efficient implementation of persistent sequences
\cite{hinze2006finger}. In this paper, we will consider the following
nested data type.

{\small\codecheck{238}
\begin{verbatim}
data Bush (a : Set) : Set where
  leaf : Bush a
  cons : a -> Bush (Bush a) -> Bush a
\end{verbatim}
}

\noindent According to Bird and Meertens \cite{bird1998nested}, the
type $\Bush\ a$ is
similar to a list where at each step down the list, entries are
\textit{bushed}. For example, a value of type $\texttt{\Bush\ \Nat}$
can be visualized as follows.
{\small\codecheck{751}
\begin{verbatim}
 bush1 = [ 4,                                        -- Nat
           [ 8, [ 5 ], [ [ 3 ] ] ],             -- Bush Nat
           [ [ 7 ], [ ], [ [ [ 7 ] ] ] ], -- Bush (Bush Nat)
           [ [ [ ], [ [ 0 ] ] ] ]   -- Bush (Bush (Bush Nat))
         ]
\end{verbatim}
}

\noindent Here, for readability, we have written
$\texttt{[x$_1$,\ldots,x$_n$]}$ instead of {\texttt{\cons\ x$_1$ (\cons\ x$_2$
  (\ldots(\cons\ x$_n$ \leaf)))}}.

Unlike ordinary data types such as lists, nested data types are
difficult to program with in total functional programming languages.
For example, in the dependently typed proof assistant Coq, the $\Bush$
data type is not definable at all, since it does not pass Coq's strict
positivity test. In Agda, $\Bush$ can be defined as a data type, but
writing functions that use this type is not trivial. For example, we
must use general recursion (rather than structural recursion) to
define the following $\hmap$ function.

{\small\codecheck{705}
\begin{verbatim}
hmap : ∀ {b c : Set} -> (b -> c) -> Bush b -> Bush c
hmap f leaf = leaf
hmap f (cons x xs) = cons (f x) (hmap (hmap f) xs)
\end{verbatim}
}
\noindent Note that, in contrast to the $\mapListprime$ function for
lists, this definition is not structurally recursive because the inner
$\hmap$ is not applied to a subterm of $\texttt{\cons\ x
  xs}$. Therefore, Agda's termination checker will reject this
definition as potentially non-terminating, unless we specify the
unsafe \texttt{--no-termination} flag.

The following function $\hfold$ for $\Bush$ is
called a \textit{higher-order fold} in the literature (e.g., \cite{bird1999generalised}, \cite{johann2007initial}).
Its definition uses $\hmap$.
{\small\codecheck{710}
\begin{verbatim}
hfold : (b : Set -> Set) ->
        (ℓ : (a : Set) -> b a) ->
        (c : (a : Set) -> a -> b (b a) -> b a) ->
        (a : Set) -> Bush a -> b a
hfold b ℓ c a leaf = ℓ a
hfold b ℓ c a (cons x xs) =
  c a x (hfold b ℓ c (b a) (hmap (hfold b ℓ c a) xs))
\end{verbatim}}

\noindent Observe that the type variable $\texttt{b}$ in $\hfold$ has
kind $\texttt{\Set\ -> \Set}$, unlike the type variable $\texttt{p}$
in $\foldList$, which has type $\Set$.
The higher-order fold $\hfold$ presents the following challenges. (1)
The definition of $\hfold$ requires the auxiliary function $\hmap$,
and $\hmap$ cannot easily be defined from $\hfold$. (2) The
definition of $\hfold$, like that of $\hmap$, is not structurally
recursive and Agda's termination checker cannot prove it to be total. (3)
Although it is possible (see below), it is fairly difficult to define
functions such as summation on $\Bush$.  (4) Unlike the induction
principle for lists, it is not clear how to obtain an induction
principle for $\Bush$ from the higher-order fold $\hfold$.

Here is the definition of a function $\ssum$ that sums up all
natural numbers in a data structure of type $\Bush\ \Nat$. Although
$\ssum$ is not a polymorphic function, it requires an auxiliary
function that is polymorphic and utilizes an argument $\texttt{k}$
that is reminiscent of continuation passing style {\cite{Plo75}}.

{\small\codecheck{718}
\begin{verbatim}
sumAux : (a : Set) -> Bush a -> (k : a -> Nat) -> Nat
sumAux =
  hfold (λ a -> (a -> Nat) -> Nat)
    (λ a k -> zero) (λ a x xs k -> add (k x) (xs (λ r -> r k)))

sum : Bush Nat -> Nat
sum ℓ = sumAux Nat ℓ (λ n -> n)
\end{verbatim}
}

\subsection{Contributions}

We present a new approach to defining fold functions for nested data
types, which we call \textit{dependently typed folds}. For
concreteness, we work within the dependently typed language
Agda. Dependently typed folds are defined by well-founded recursion,
hence their termination is easily confirmed by Agda. Map functions and
many other terminating functions can be defined directly from the
dependently typed folds. Moreover, the higher-order folds (such as
$\hfold$) are definable from the dependently typed folds. In addition,
the definitions of dependently typed folds can easily be generalized
to corresponding induction principles. Thus we can formally reason
about programs involving nested data types in a total dependently
typed language.  While we illustrate these ideas by focusing on the
$\Bush$ example, our approach also works for other kinds of nested
data types; see Section~\ref{sec:more-discussion} for an example.

\section{Dependently typed fold for Bush}
\label{sec:nfold}

Let us continue the consideration of the $\Bush$ data type. The
following is the result of evaluating $\texttt{\hmap\ f \bushone}$,
where $\bushone$ is the data structure defined in the introduction,
and $\texttt{f : \Nat\ -> b}$ for some type $\texttt{b}$.

{\small\codecheck{752}
\begin{verbatim}
  [ f 4,                                            -- b
    [ f 8, [ f 5 ], [ [ f 3 ] ] ],             -- Bush b
    [ [ f 7 ], [ ], [ [ [ f 7 ] ] ] ],  -- Bush (Bush b)
    [ [ [ ], [ [ f 0 ] ] ] ]     -- Bush (Bush (Bush b))
  ]
\end{verbatim}
}
\noindent To motivate the definition of the dependently typed fold
below, we first consider the simpler question of how to define a map
function for $\Bush$ by structural recursion. The reason our
definition of $\hmap$ in the introduction was not structural is that
in order to define the map function for $\texttt{\Bush\ \Nat}$, we
need to already have the map functions defined for $\Bush^n\ \Nat =
\Bush\ (\Bush\ (\ldots (\Bush\ \Nat)))$ for
all $n \geq 0$, which seems paradoxical.  Our solution is to define a
general map function for $\Bush^n$, for all $n \geq 0$.  First we
define a type-level function $\NTimes$ such that $\texttt{\NTimes\ n
  b} = \texttt{b}^n$:
{\small\codecheck{185}
\begin{verbatim}
NTimes : (n : Nat) -> (b : Set -> Set) -> Set -> Set
NTimes zero b a = a
NTimes (succ n) b a = b (NTimes n b a)

\end{verbatim}}

\noindent We can now define the following
map function for $\Bush^n$:

{\small\codecheck{726}
\begin{verbatim}
nmap : ∀ {a b : Set} -> (n : Nat) -> (a -> b) ->
       NTimes n Bush a -> NTimes n Bush b
nmap zero f x = f x
nmap (succ n) f leaf = leaf
nmap (succ n) f (cons x xs) =
  cons (nmap n f x) (nmap (succ (succ n)) f xs)
\end{verbatim}
}
\noindent Note that $\texttt{\nmap\ 1}$ corresponds to the map
function for $\texttt{\Bush\ a}$.  The recursive definition of $\nmap$
is well-founded because all the recursive calls are on the components
of the constructor $\cons$. The Agda termination checker accepts this
definition of $\nmap$.

We are now ready to introduce the dependently typed fold. The idea is
to define the fold over the type $\texttt{\NTimes\ n \Bush}$
simultaneously for all $n$.
{\small\codecheck{245}
\begin{verbatim}
nfold : (p : Nat -> Set) ->
        (ℓ : (n : Nat) -> p (succ n)) ->
        (c : (n : Nat) -> p n -> p (succ (succ n)) -> p (succ n)) ->
        (a : Set) -> (z : a -> p zero) ->
        (n : Nat) -> NTimes n Bush a -> p n
nfold p ℓ c a z zero x = z x
nfold p ℓ c a z (succ n) leaf = ℓ n
nfold p ℓ c a z (succ n) (cons x xs) =
  c n (nfold p ℓ c a z n x) (nfold p ℓ c a z (succ (succ n)) xs)
\end{verbatim}
}

The dependently typed fold $\nfold$ captures the most general form of
computing/traversal on the type $\texttt{\NTimes\ n
  \Bush\ a}$. Similarly to $\nmap$, the definition of $\nfold$ is
well-founded. Note that unlike the $\hfold$ in the introduction, this
definition of fold does not require a map function to be defined
first. In fact, $\nmap$ is definable from $\nfold$:
{\small\codecheck{284}
\begin{verbatim}
nmap : ∀ {a b : Set} -> (n : Nat) -> (a -> b) ->
       NTimes n Bush a -> NTimes n Bush b
nmap {a} {b} n f ℓ =
  nfold (λ n -> NTimes n Bush b) (λ n -> leaf) (λ n -> cons) a f n ℓ
\end{verbatim}}

We can also prove that $\texttt{\nmap\ 1}$ satisfies the defining
properties of $\hmap$ from the introduction. Let $\hmapprime =
\texttt{\nmap\ 1}$.

\pagebreak

{\small\codecheck{421}
\begin{verbatim}
lemma-nmap : ∀ {a b : Set} -> (f : a -> b) -> (m n : Nat) ->
             (x : NTimes (add m n) Bush a) ->
             nmap (add m n) f x == nmap m (nmap n f) x
lemma-nmap f zero n x = refl
lemma-nmap f (succ m) n leaf = refl
lemma-nmap f (succ m) n (cons x xs) =
  cong2 cons (lemma-nmap f m n x) (lemma-nmap f (succ (succ m)) n xs)

hmap-leaf : ∀ {a b : Set} -> (f : a -> b) -> hmap' f leaf == leaf
hmap-leaf f = refl

hmap-cons : ∀ {a b : Set} -> (f : a -> b) -> (x : a) ->
            (xs : Bush (Bush a)) ->
            hmap' f (cons x xs) == cons (f x) (hmap' (hmap' f) xs)
hmap-cons f x xs = cong (cons (f x)) (lemma-nmap f 1 1 xs)
\end{verbatim}
}

Many other terminating functions can also be conveniently defined in
term of $\nfold$. For example, the summation of all the entries in
$\texttt{\Bush\ \Nat}$ and the length function for $\texttt{\Bush}$
can be defined as follows:

{\small\codecheck{294}
\begin{verbatim}
sum : Bush Nat -> Nat
sum =
  nfold (λ n -> Nat) (λ n -> zero) (λ n -> add) Nat (λ x -> x) 1

length : (a : Set) -> Bush a -> Nat
length a =
  nfold (λ n -> Nat) (λ n -> zero) (λ n r1 r2 -> succ r2)
    a (λ x -> zero) 1
\end{verbatim}
}

Note that this definition of $\ssum$ is much more natural and
straightforward than the one we gave in the introduction.

\section{Induction principle for Bush}

While there is no obvious induction principle corresponding to the
higher-order fold $\hfold$, we can easily generalize the dependently
typed fold $\nfold$ to obtain an induction principle for $\Bush$. The
following function $\ind$ is related to $\nfold$ in the same way that
the induction principle for $\List$ is related to its fold function.

{\small\codecheck{306}
\begin{verbatim}
ind : ∀ {a : Set} -> {p : (n : Nat) -> NTimes n Bush a -> Set} ->
      (base : (x : a) -> p zero x) ->
      (ℓ : (n : Nat) -> p (succ n) leaf) ->
      (c : (n : Nat) -> (x : NTimes n Bush a) ->
        (xs : NTimes (succ (succ n)) Bush a) ->
        p n x -> p (succ (succ n)) xs -> p (succ n) (cons x xs)) ->
      (n : Nat) -> (xs : NTimes n Bush a) -> p n xs
ind base ℓ c zero xs = base xs
ind base ℓ c (succ n) leaf = ℓ n
ind base ℓ c (succ n) (cons x xs) =
  c n x xs (ind base ℓ c n x) (ind base ℓ c (succ (succ n)) xs)
\end{verbatim}}

Observe that $\ind$ follows the same structure as $\nfold$. The type
variable $\texttt{p}$ is generalized to a predicate of kind
$\texttt{(n : \Nat) -> \NTimes\ n \Bush\ a -> \Set}$. The type of $\ind$
specifies how to prove by induction that a property $\texttt{p}$ holds
for all members of the type $\texttt{\NTimes\ n \Bush\ a}$.  More
specifically, for the base case, we must show that $\texttt{p}$ holds
for any $\texttt{x}$ of type $\texttt{\NTimes\ \zero\ \Bush\ a}$
(which equals $\texttt{a}$), hence $\texttt{p \zero\ x}$. For the leaf
case, we must show that $\texttt{p}$ holds for $\leaf$ of type
$\texttt{\NTimes\ (\suc\ n) \Bush\ a}$. For the cons case, we assume
as the induction hypotheses that $\texttt{p}$ holds for some
$\texttt{x}$ of type $\texttt{\NTimes\ n \Bush\ a}$ and some
$\texttt{xs}$ of type $\texttt{\NTimes\ (\suc\ (\suc\ n)) \Bush\ a}$,
and then we must show that $\texttt{p}$ holds for $\texttt{\cons\ x
  xs}$.

With $\ind$, we can now prove properties of $\nmap$. For example, the
following is a proof that $\nmap$ has the usual identity property of
functors.

{\small\codecheck{345}
\begin{verbatim}
nmap-id : ∀ {a : Set} -> (n : Nat) -> (y : NTimes n Bush a) ->
          nmap n (id a) y == y
nmap-id {a} n y =
  ind {a} {λ n xs -> nmap n (id a) xs == xs} (λ x -> refl) (λ n -> refl)
    (λ n x xs ih1 ih2 -> cong2 cons ih1 ih2) n y
\end{verbatim}
}

\noindent We note that the usual way of proving things in Agda is by
recursion, relying on the Agda termination checker to prove
termination. Our purpose here, of course, is to illustrate that our
induction principle is strong enough to prove many properties without
needing Agda's recursion. Nevertheless, the above proof is equivalent
to the following proof by well-founded recursion.

{\small\codecheck{352}
\begin{verbatim}
nmap-id' : ∀ {a : Set} -> (n : Nat) -> (y : NTimes n Bush a) ->
           nmap n (id a) y == y
nmap-id' zero y = refl
nmap-id' (succ n) leaf = refl
nmap-id' (succ n) (cons x y) =
  cong2 cons (nmap-id' n x) (nmap-id' (succ (succ n)) y)
\end{verbatim}
}

\noindent The first two clauses of $\texttt{nmap-id'}$ correspond
to the two arguments \texttt{(λ n -> refl)} for
$\texttt{nmap-id}$. The recursive calls $\texttt{nmap-id' n x}$ and
\texttt{nmap-id' (\suc\ (\suc\ n)) y} in the definition of
$\texttt{nmap-id'}$ correspond to the inductive hypotheses
$\texttt{ih1}$ and $\texttt{ih2}$ in $\texttt{nmap-id}$.

\section{Higher-order folds and dependently typed folds}

Comparing $\nfold$, the dependently typed fold that was defined in
Section~\ref{sec:nfold}, to $\hfold$, the higher-order fold defined
in the introduction, we saw that $\nfold$ does not depend on $\nmap$,
and $\nmap$ can be defined from $\nfold$. We also saw that the
termination of $\nfold$ is obvious and that it can be used to define
other terminating functions.

In this section, we will show the $\hfold$ is actually equivalent to
$\nfold$ in the sense that they are definable from each other.

\subsection{Defining $\hfold$ from $\nfold$}

Using $\nfold$, it is straightforward to define $\hfold$, because the
latter is essentially the former instantiated to the case $n=1$.

{\small\codecheck{256}
\begin{verbatim}
hfold : (b : Set -> Set) ->
        (ℓ : (a : Set) -> b a) ->
        (c : (a : Set) -> a -> b (b a) -> b a) ->
        (a : Set) -> Bush a -> b a
hfold b ℓ c a x =
  nfold (λ n -> NTimes n b a) (λ n -> ℓ (NTimes n b a))
    (λ n -> c (NTimes n b a)) a (λ x -> x) 1 x
\end{verbatim}
}

We can prove that this version of $\hfold$ satisfies the defining
properties of the version of $\hfold$ that was defined in the
introduction (and therefore the two definitions agree).  Since the
proof of $\texttt{hfold-cons}$ is rather long, we have omitted it, but
the full machine-checkable proof can be found at {\cite{depfold-agda}}.

{\small\codecheck{440}
\begin{verbatim}
hfold-leaf : (a : Set) -> (p : Set -> Set) ->
              (ℓ : (b : Set) -> p b) ->
              (c : (b : Set) -> b -> p (p b) -> p b) ->
               hfold p ℓ c a leaf == ℓ a
hfold-leaf a p ℓ c = refl

hfold-cons : (a : Set) -> (p : Set -> Set) ->
              (ℓ : (b : Set) -> p b) ->
              (c : (b : Set) -> b -> p (p b) -> p b) ->
              (x : a) -> (xs : Bush (Bush a)) ->
              hfold p ℓ c a (cons x xs)
              == c a x (hfold p ℓ c (p a) (hmap (hfold p ℓ c a) xs))
hfold-cons a p ℓ c x xs = ...
\end{verbatim}
}

\subsection{Defining $\nfold$ from $\hfold$}

The other direction is much trickier. In attempting to define $\nfold$
from $\hfold$, the main difficulty is that we must supply a type
function $\texttt{b : \Set\ -> \Set}$ to $\hfold$, and this
$\texttt{b}$ should somehow capture the quantification over natural
numbers. Ideally, we would like to define $\texttt{b}$ such that
$\texttt{b}^n\,\texttt{a} = \texttt{p}\,n$ for all $n$ and some
suitable $\texttt{a}$. However, this is clearly impossible, because
$\texttt{p}$ is an arbitrary type family, which can be defined so that
$\texttt{p\,0} = \texttt{p\,1}$ but $\texttt{p\,1}\neq
\texttt{p\,2}$. This would imply $\texttt{a} = \texttt{b}\,\texttt{a}$
but $\texttt{b}\, \texttt{a} \neq \texttt{b}^2\,\texttt{a}$, a
contradiction.

Surprisingly, it is possible to work around this by arranging things
so that there is a canonical function $\texttt{b}^n\,\texttt{a} \to
\texttt{p}\,n$, rather than an equality. This is done by defining the
following rather unintuitive type-level function $\PS$.
{\small\codecheck{216}
\begin{verbatim}
PS : (p : Nat -> Set) -> Set -> Set
PS p A = (n : Nat) -> (A -> p n) -> p (succ n)
\end{verbatim}
}
\noindent The type $\texttt{\PS\ p}$ is special because there is a map
$\texttt{\NTimes\ n (\PS\ p) a} \to \texttt{p n}$.

{\small\codecheck{220}
\begin{verbatim}
PS-to-P : (p : Nat -> Set) -> (a : Set) -> (z : a -> p zero) ->
          (n : Nat) -> NTimes n (PS p) a -> p n
PS-to-P p a z zero x = z x
PS-to-P p a z (succ n) hyp = hyp n ih
  where
    ih : NTimes n (PS p) a -> p n
    ih = PS-to-P p a z n
\end{verbatim}
}

\noindent So if we set
$\texttt{b} = \texttt{\PS\,p}$, we have the promised canonical map
$\texttt{b}^n\,\texttt{a} \to \texttt{p}\,n$. We can pass this
$\texttt{b}$ to $\hfold$ to go from $\texttt{\Bush\ a}$ to
$\texttt{\PS\ p a}$.
{\small\codecheck{470}
\begin{verbatim}
fold-PS : (p : Nat -> Set) ->
         (ℓ : (n : Nat) -> p (succ n)) ->
         (c : (n : Nat) -> p n -> p (succ (succ n)) -> p (succ n)) ->
         (a : Set) -> Bush a -> PS p a
fold-PS p ℓ c =
  hfold (PS p) (λ a n tr -> ℓ n)
    (λ a x xs n tr -> c n (tr x) (xs (succ n) (λ f -> f n tr)))
\end{verbatim}
}

\noindent Now, provided that we are able to \textit{lift} the
function $\texttt{\Bush\ a -> \PS\ p a}$ to its $n$th iteration, i.e.,
to a function of type 
$\texttt{\NTimes\ n \Bush\ a -> \NTimes\ n (\PS\ p) a}$, then we will
be able to define the dependently typed fold via the following.

\pagebreak

{\small\codecheck{479}
\begin{verbatim}
nfold' : (p : Nat -> Set) ->
         (ℓ : (n : Nat) -> p (succ n)) ->
         (c : (n : Nat) -> p n -> p (succ (succ n)) -> p (succ n)) ->
         (a : Set) -> (z : a -> p zero) ->
         (n : Nat) -> NTimes n Bush a -> p n
nfold' p ℓ c a z n x = PS-to-P p a z n (lift n x)
  where
    lift : (n : Nat) -> NTimes n Bush a -> NTimes n (PS p) a
    lift n x =
      liftNTimes Bush (PS p) (λ a b -> hmap) n (fold-PS p ℓ c) a x

\end{verbatim}}
\noindent The $\liftNTimes$ function can indeed be defined by induction
on natural numbers.

{\small\codecheck{190}
\begin{verbatim}
liftNTimes : (b c : Set -> Set) ->
             (∀ x y -> (x -> y) -> (b x -> b y)) ->
             (n : Nat) -> (∀ a -> b a -> c a) ->
             (a : Set) -> NTimes n b a -> NTimes n c a
liftNTimes b c m zero f a x = x
liftNTimes b c m (succ n) f a x =
  f (NTimes n c a)
    (m (NTimes n b a) (NTimes n c a) (liftNTimes b c m n f a) x)
\end{verbatim}}
Finally, we can prove that the function $\nfoldprime$ that we just
defined behaves identically to the $\nfold$ that was defined in
Section~\ref{sec:nfold}. Again, since the proof is rather long and
uses several lemmas, we do not reproduce it here. The machine-checkable
proof can be found at {\cite{depfold-agda}}.

{\small\codecheck{673}
\begin{verbatim}
theorem : ∀ p ℓ c a z n x ->
          nfold p ℓ c a z n x == nfold' p ℓ c a z n x
theorem p ℓ c a z n x = ...
\end{verbatim}
}

\section{Nested data types beyond Bush}
\label{sec:more-discussion}

So far, we have focused on the $\Bush$ type, but our approach works
for arbitrary nested data types, including ones that are defined by
mutual recursion. To illustrate this, consider the following pair of
mutually recursive data types:

{\small\codecheck{714}
\begin{verbatim}
data Bob (a : Set) : Set
data Dylan (a b : Set) : Set

data Bob a where
  robert : a -> Bob a
  zimmerman : Dylan (Bob (Dylan a (Bob a))) (Bob a) -> Bob (Dylan a a) -> Bob a

data Dylan a b where
  duluth : Bob a -> Bob b -> Dylan a b
  minnesota : Dylan (Bob a) (Bob b) -> Dylan a b
\end{verbatim}
}

\noindent
As usual, the higher-order fold is easy to define. There are two
separate such folds, one for $\Bob$ and one for $\Dylan$:

\pagebreak

{\small\codecheck{X}
\begin{verbatim}
hfold-bob : (bob : Set -> Set) ->
      (dylan : Set -> Set -> Set) ->
      (rob : ∀ a -> a -> bob a) ->
      (zim : ∀ a -> dylan (bob (dylan a (bob a))) (bob a) -> bob (dylan a a) -> bob a) ->
      (dul : ∀ a b -> bob a -> bob b -> dylan a b) ->
      (min : ∀ a b -> dylan (bob a) (bob b) -> dylan a b) ->
      ∀ a -> Bob a -> bob a

hfold-dylan : (bob : Set -> Set) ->
      (dylan : Set -> Set -> Set) ->
      (rob : ∀ a -> a -> bob a) ->
      (zim : ∀ a -> dylan (bob (dylan a (bob a))) (bob a) -> bob (dylan a a) -> bob a) ->
      (dul : ∀ a b -> bob a -> bob b -> dylan a b) ->
      (min : ∀ a b -> dylan (bob a) (bob b) -> dylan a b) ->
      ∀ a b -> Dylan a b -> dylan a b
\end{verbatim}
}
  
The dependent fold requires some explanation. Recall that for $\Bush$,
the only type expressions of interest were of the form
$\Bush^n\,\texttt{a}$, so we used the natural number $n$ to index
these types. In the more general case, we must consider more
complicated type expressions such as
$\Dylan\,(\Bob\,\texttt{a})\,(\Dylan\,\texttt{a}\,\texttt{b})$.
Therefore, we need to replace the natural numbers with a custom
type. We define a type $\BobDylanIndex$, which represents expressions
built up from type variables and the type constructors $\Bob$ and
$\Dylan$.

{\small\codecheck{755}
\begin{verbatim}
data BobDylanIndex : Set where
  varA : BobDylanIndex
  varB : BobDylanIndex
  BobC : BobDylanIndex -> BobDylanIndex
  DylanC : BobDylanIndex -> BobDylanIndex -> BobDylanIndex
\end{verbatim}
}

We can then give an interpretation function for these type
expressions. This plays the role that $\NTimes$ played in the $\Bush$
case:

{\small\codecheck{764}
\begin{verbatim}
I : (Set -> Set) -> (Set -> Set -> Set) -> Set -> Set -> BobDylanIndex -> Set
I bob dylan a b varA = a
I bob dylan a b varB = b
I bob dylan a b (BobC expr) = bob (I bob dylan a b expr)
I bob dylan a b (DylanC expr1 expr2) = dylan (I bob dylan a b expr1) (I bob dylan a b expr2)
\end{verbatim}
}

\noindent
For example, if
\[\texttt{\small i = DylanC (BobC varA) (DylanC varA varB)},
\]
then
\[\texttt{\small I bob dylan a b i = dylan (bob a) (dylan a b)}.
\]
The dependent fold is defined simultaneously for $\Bob$ and $\Dylan$,
and in fact for all type expressions that are built from $\Bob$ and
$\Dylan$. Its type is the following:

{\small\codecheck{X}
\begin{verbatim}
nfold : (p : BobDylanIndex -> Set) ->
        (rob : ∀ a -> p a -> p (BobC a)) ->
        (zim : ∀ a -> p (DylanC (BobC (DylanC a (BobC a))) (BobC a))
                      -> p (BobC (DylanC a a)) -> p (BobC a)) ->
        (dul : ∀ a b -> p (BobC a) -> p (BobC b) -> p (DylanC a b)) ->
        (min : ∀ a b -> p (DylanC (BobC a) (BobC b)) -> p (DylanC a b)) ->
        (a b : Set) ->
        (baseA : a -> p varA) ->
        (baseB : b -> p varB) -> 
        (∀ i -> I Bob Dylan a b i -> p i)
\end{verbatim}
}

Note that although the types $\Bob$ and $\Dylan$ are complicated, the
corresponding $\nfold$ can be systematically derived from their
definition. Moreover, as in the case of $\Bush$, the higher-order
folds and the dependent fold are definable in terms of each other. In
addition, the induction principle, which generalizes $\nfold$, can be
easily defined. Full details can be found in the accompanying code
{\cite{depfold-agda}}.

\section{Discussion}

We think that the equivalence of $\hfold$ and $\nfold$ is both
surprising and useful. The reason it is surprising is because it was
informally believed among researchers that $\hfold$ is too abstract
for most useful programming tasks. The reason it is potentially useful
is that in the context of some dependently typed programming languages
or proof assistants (such as Coq), when the user writes a data type
declaration, the system should automatically derive the appropriate
folds and induction principles for the data type. In the case of
nested data types, there is currently no universally good way to do
this (which is presumably one of the reasons Coq does not support the
$\Bush$ type). Now on the one hand, we have $\nfold$, which is a
practical programming primitive, but its type is not easy to generate
from a user-defined data type declaration. For example, even stating
the type of $\nfold$ requires a reference to an ancillary data type,
which is $\Nat$ in the case of $\Bush$ but can be more complicated for
a general nested type. On the other hand, we have $\hfold$, which is
not very practical, but its type can be easily read off from a data
type declaration. The fact that we have shown $\nfold$ to be definable
in terms of $\hfold$ suggests a solution to this problem: given a data
type declaration, the system can generate its corresponding $\hfold$,
and then the user can follow a generic recipe to derive the more
useful $\nfold$.

\section{Conclusion and future work}

Using $\Bush$ as an example, we showed how to define dependently typed
folds for nested data types.  Unlike higher-order folds, dependently
typed folds can be used to define maps and other terminating
functions, and they have analogous induction principles, similar to
the folds for ordinary data types. We showed how to reason about
programs involving nested data types in Agda. Last but not least, we
also showed that dependently typed folds and higher-order folds are
mutually definable. This has some potential applications in
implementations of dependent type theories, because given a
user-defined nested data type, the corresponding higher-order fold can
be automatically generated, and then the user can derive the more
useful dependent fold by following a generic recipe. All of our proofs
are done in Agda, without using any unsafe flag.

Our long term goal is to derive induction principles for \textit{any}
algebraic data type (nested or non-nested). There is still a lot of
work to be done. In this paper, we only showed how to get the
dependently typed fold and induction principle for the single example
of $\Bush$. Although our approach also works for other nested data
types, we have not yet given a formal characterization of dependently
typed folds and their induction principles in the general
case. Another research direction is to study the direct relationship
between the induction principles (derived from dependently typed
folds) and higher-order folds. In the $\Bush$ example, it corresponds
to asking if we can define $\ind$ from $\hfold$, possibly with some
extra properties that can also be read off from the data type definition.

\section*{Acknowledgements}

We thank the referees for their thoughtful comments. This work was
supported by the Natural Sciences and Engineering Research Council of
Canada (NSERC) and by the Air Force Office of Scientific Research
under Award No.\@ FA9550-21-1-0041.

\pagebreak

\bibliography{dep}

\end{document}